
\input phyzzx

\pubnum{92-58}
\titlepage
\title{TEVATRON MASS LIMITS FOR HEAVY QUARKS DECAYING VIA FLAVOR CHANGING
NEUTRAL CURRENT}
\author{Biswarup Mukhopadhyaya\foot{Address from November 1, 1992: Mehta
Research Institute, 10, Kasturba Gandhi Marg, Allahabad 211 002, INDIA}
nd D.P. Roy}
\address{Theoretical Physics Group\break Tata Institute of Fundamental
Research, Homi Bhabha Road, Bombay 400 005, INDIA}
\abstract{The dimuon and dielectron data from the Tevatron $\bar pp$
collider are used to probe for heavy quarks, which decay dominantly via
flavour changing neutral current. Depending on whether the $FCNC$ decay
occurs at the tree or loop level, one gets a lower mass limit of 85 or 75
GeV. The former applies to singlet, vector doublet and mirror type quarks
while the latter applies to a lefthanded quark doublet of the fourth
generation.}

\endpage

By far the most extensive search for the top quark has been carried out by
the CDF experiment [1] at the Tevatron $\bar pp$ collider, leading to a
95\% CL mass limit of $m_t > 91$ GeV. The most important channels for this
top search program are the isolated large-$p_T$ dilepton channels $e\mu,
\mu\mu$ and $ee$, which account for a mass bound of $m_t > 86$ GeV. The
relevant processes for these channels are $t\bar t$ pair production by
gluon-gluon and quark-antiquark fusion
$$
gg(q\bar q) \rightarrow t\bar t
\eqno (1)
$$
followed by the charged current semileptonic decay of both the top quarks
$$
t \rightarrow b \ell^+\nu, ~~~\bar t \rightarrow \bar b \ell^- \bar \nu
\eqno (2)
$$
where $\ell = e, \mu$. With the $e$ and $\mu$ decay branching ratios of
10\% each one gets an overall branching ratio of 4\% for $t\bar t$ decay
into the above dilepton channels. Together with the dilepton detection
efficiency of about 16\%, arising from the various cuts, this results in
an overall efficiency of $t\bar t$ detection in the dilepton channels
$\simeq 0.6$\%.

It has been generally recognised that one has the same dilepton branching
ratio and a similar detection efficiency for any other heavy quark
decaying via charged current interaction [2].  Combined with the QCD
prediction of same production cross-section, it implies that the above
mentioned mass limit of 86 GeV holds for all such quarks. However, the
first set of conditions would not apply to quarks, which decay mainly via
flavour changing neutral current. Indeed, there has been no search for
such heavy quarks using the Tevatron data so far. The present work is
devoted to this excercise. We shall see below that a systematic analysis
of the Tevatron $\mu\mu$ and $ee$ data leads to an equally strong mass
limit for these quarks as well. More precisely one gets a lower mass limit
of 85 or 75 GeV depending on whether the $FCNC$ decay occurs at the tree or
the loop level.

Let us first recall the models, where one may expect a heavy quark $Q$ to
decay dominantly via $FCNC$. For singlet, vector doublet and mirror type
quarks [3] both $FCNC$ and $CC$ decays occur at the tree level and are
generally of comparable magnitude. Now consider a singlet quark of charge
$-1/3$, which occurs e.g. in the $E_6$ grand unified model [4]. If this
quark is lighter than the top then only the $FCNC$ decay into the 3rd
generation would be kinematically allowed. In this case one expects the
$FCNC$ decay to dominate, provided the ratio of the mixing angles are in
the range
$$
\sin^2 \theta_{Qs,Qd}/\sin^2 \theta_{Qb} \lsim 10^{-1}.
\eqno (3)
$$
According to most of the fermion mass matrix models [3]
$$
\sin^2 \theta_{Qi} \sim (m_i/m_Q)^{1~{\rm to}~2}
\eqno (4)
$$
implying
$$
\sin^2 \theta_{Qs}/\sin^2 \theta_{Qb} \sim (m_s/m_b)^{1~{\rm to}~2} \sim
10^{-1 ~{\rm to}~ -2}.
\eqno (5)
$$
The corresponding ratio for $d$ quark is of course much smaller. Thus for
the plausible range of the mixing angles one expects dominant $FCNC$ decay.
The same holds true for a charge $-1/3$ quark of a vector doublet [5] or a
mirror doublet [6], provided it is lighter than its accompanying charge $2/3$
quark as well as the top. The above three models shall be collectively
referred to as exotic quark models, since they correspond to non-standard
$SU(2)$ representations.

Consider next a charge $-1/3$ quark belonging to 4th generation of the standard
$SU(2)$ doublet. In this case the $FCNC$ decay occurs only at the one-loop
level due to the GIM cancellation. So $FCNC$ dominance holds only over a
limited range of the ratio (3), i.e.
$$
\sin^2 \theta_{Qs}/\sin^2 \theta_{Qb} \lsim 10^{-4}.
\eqno (6)
$$
This is evidently outside the range (5). On the other hand it may be
reasonable in models where one approximates the $4 \times 4$ $CKM$ matrix by a
block diagonal form containing two $2 \times 2$ matrices [7-9]. This
corresponds to a scenario where the mixing between the first two
generations, as well as that between the last two, are substantial, whereas
the two pairs of generations mix rather feebly with each other. In any
case the range (6) is a phenomenologically allowed part of the parameter space
[3,10]. Hence it is necessary to extend the heavy quark search to the $FCNC$
decay channel in order to close this window.

Thus there is a good chance of dominant $FCNC$ decay for the charge -1/3
quark of the exotic models and (to a lesser extent) for that of the 4th
generation. Hence it is important to search for such quarks in the $FCNC$
decay channels. Besides, the search for a possible heavy quark with
dominant $FCNC$ decay is important for another reason -- i.e. its nuissance
value to the Higgs and $SUSY$ search programmes at hadron colliders, as
emphasised in refs [8,9]. For, a heavy quark pair decaying via $FCNC$ will
be a formidable source of 4 lepton $(ZZ \rightarrow \ell^+\ell^- \ell^+
\ell^-)$ and missing-$p_T$ $(ZZ \rightarrow \nu\bar\nu \nu\bar\nu)$
events, which are crucial channels for Higgs and $SUSY$ searches
respectively.

We should mention here that the current mass limits for quarks with
dominant $FCNC$ decay come from $LEP$ [11], i.e.
$$
m_Q > M_Z/2.
\eqno (7)
$$
Therefore we shall restrict our search to above this region.

Our analysis is based on a parton level Monte Carlo calculation.  The
$Q\bar Q$ production process is identical to that of $t\bar t$ production,
as given by eq. (1).  The results presented below have been obtained with
the structure functions and the $QCD$ coupling parameter of $GHR$ [12].
The more recent parametrisations give a somewhat smaller cross-section due
to a smaller value of the $QCD$ coupling parameter $\wedge$.  We have
checked this using the parametrisation of $DFLM$ [13], which reduces the
size of the cross-section by $\sim 25$\%.  On the other hand the higher
order $QCD$ effects, not included in this calculation, are expected to
enhance the cross-section by $\sim 50$\% [14].  Thus we expect that the
uncertainty in the $QCD$ parametrisation and the higher order $QCD$
effects can reduce or enhance the size of the cross-section presented
below by a factor of 1.5.

For the exotic quarks, we have tree level $FCNC$ decay which is easy to
handle.  Since the tree level $FCNC$ decay proceeds necessarily via $Z$,
the decay channel of our interest
$$
Q(p) \rightarrow b(p') \ell^+(p_1) \ell^-(p_2)
\eqno (8)
$$
has a brnaching ratio of 3.3\% for each lepton species.  For a singlet $Q$
the tree level $FCNC$ arises from the mixing of lefthanded quarks.  The
resulting matrix element for the above decay is
$$
M = {g^2 \sin \theta_{Qb} \cos \theta_{Qb} \over 2\cos^2 \theta_W} \cdot
{\bar u_b (p') \gamma_\mu \left[{1 - \gamma_5 \over 2}\right] u_Q(p) \bar
u_\ell (p_1) \gamma_\mu\left[\alpha {1 - \gamma_5 \over 2} + \beta {1 +
\gamma_5 \over 2}\right] v_\ell (p_2) \over (p_1 + p_2)^2 - m^2_Z}
\eqno (9)
$$
where
$$
\alpha = -1/2 + \sin^2 \theta_W, ~~~\beta = \sin^2 \theta_W
\eqno (10)
$$
The corresponding squared matrix element is
$$
|M|^2 = {g^4 \sin^2 \theta_{Qb} \cos^2 \theta_{Qb} \left[\beta^2 (m^2_Q -
s_2) (s_2 - m^2_b) + \alpha^2(m^2_Q - s_3) (s_3 - m^2_b)\right] \over
2\cos^4 \theta_W \left[(s_1 - m^2_Z)^2 + m^2_Z \Gamma^2_Z\right]}
\eqno (11)
$$
$$
s_1 = (p - p')^2, ~~~s_{2,3} = (p - p_{1,2})^2.
$$
The quark mixing angle drops out from the resulting distribution function
$d\Gamma/\Gamma$, where $\Gamma$ is the partial decay width for (8). The
dilepton differencial cross-section is obtained by convoluting this
quantity with the $Q\bar Q$ production cross-section together with the
above branching ratio. For a vector doublet $Q$ the $FCNC$ arises from the
mixing of right landed quarks. The corresponding squared matrix element is
obtained from (11) simply by interchanging $\alpha^2 ~{\rm and}~ \beta^2$. For
a mirror doublet the $FCNC$ arises from the mixing of both lefthanded and
righthanded quarks. In this case the quark mixing-angles would not in
general drop out of $d\Gamma/\Gamma$. It is well known however that this
quantity is insensitive to the squared matrix element and depends mainly
on the phase space factor. Thus for simplicity we assume equal mixing
angles for the left and right handed quarks. In this case the mixing angle
factors out and the squared matrix element corresponds to (11) with
$\alpha^2$ and  $\beta^2$ each replaced by $(\alpha^2 + \beta^2)$. The
resulting dilepton cross-sections are practically identical for the three
cases, because (1) the decay distributions are insensitive to the squared
matrix element as mentioned above and (2) $\alpha^2 \simeq \beta^2$ for
$\sin^2\theta_W \simeq  0.23$. Therefore, they shall be represented below by a
common set of curves, obtained with eq. (11).

For the 4th generation quark the one-loop $FCNC$ decay proceeds via $Z$ as
well as $\gamma$ and the gluon. Consequently the branching ratio for the
decay (8) is lower in this case. The exact value depends on the masses of
$t$ and the corresponding 4th generation quark $t'$, which appear as
internal lines in the loop diagrams. We shall conservatively assume this
branching ratio to be = 1\%, which corresponds to $m_t = 100$ and $m_{t'}
= 200$ GeV [8]. As one can see from the last paper of ref [7], this
branching ratio increases with increasing $m_t$ or $m_{t'}$ -- increasing
by a factor of 2 for $m_t \rightarrow 120$ GeV and a factor of 3 for
$m_{t'} \rightarrow 300$ GeV.   The former
would compensate for any decrease coming from $m_{t'}$ being less than 200
GeV. It should be noted here that under our assumption of negligible
mixing between the fourth generation and the first two, the mixing between
the third and fourth generations can be parametrised by a single angle. It
appears as a common factor in all the loop diagrams mentioned above and
hence drops out from the branching ratio. For the same reason it also
drops out from the distribution function $d\Gamma/\Gamma$ for the decay
process (8).  To obtain this distribution function one has to compute the
contributing one-loop diagrams. In the 't Hooft-Feynman gauge, there are
10 such diagrams; 6 of these entail computation of three-point functions
and the remaining 4 of two-point functions. Since there are massive
particles in the external legs, we have to make use of the algorithm
involving form-factors that are complicated functions of the external and
internal masses. Ultimately all these form factors are expressible in
terms of Spence functions [15]. The effective $QbZ$ vertex turns out to be
of the form
$$
{\cal L}^{eff}_{QbZ} = \bar u_b(p')\left[\rho\gamma_\mu {(1-\gamma_5)
\over 2} + \lambda p_\mu {(1-\gamma_5) \over 2} + \eta p'_\mu
{(1+\gamma_5) \over 2} \right] u_Q(p)
\eqno (12)
$$
where $\rho, \lambda$ and $\eta$ can be written in terms of the above
mentioned form factors. The expressions for these quantities and the
resulting squared matrix element are too long to write down here [16].
Some of these quantities may be found in ref. [7]. Although the squared
matrix element depends on the masses of $t$ and $t'$, the dependence in
rather smooth. Consequently the resulting distribution function
$d\Gamma/\Gamma$ is very insensitive to these masses.

Before presenting the results let us summarise the main features of the
$CDF$ $\mu\mu$ and $ee$ data [1], which are relevant to our analysis. The
data sample corresponds to an integrated luminosity of $4.1~ pb^{-1}$ for
electrons and $3.5~ pb^{-1}$ for muons, taken at a $CM$ energy of 1.8 TeV.
The relevant cuts and efficiency factors are as follows:
\item{i)} Each lepton has a $p_T$ cut of
$$
p^\ell_T > 15 ~{\rm GeV}.
$$

\item{ii)} Muons are detected by the central muon detector and by minimum
ionisation over the rapidity intervals
$$
{\eta_\mu} = 0 - 0.6 (CM) ~{\rm and}~ 0.6 - 1.2 (MI),
$$
while the electrons are detected by the central and plug electromagnetic
calorimeters over
$$
{\eta_e} = 0 - 1 (CE) ~{\rm and}~ 1.26 - 2.22 (PE)
$$

\item{iii)} There is an isolation cut on each lepton requiring the accompanying
$E_T$ within a cone of radius $\Delta R = [\Delta \phi^2 + \Delta
\eta^2]^{1/2} = 0.4$ to be
$$
E_T (\Delta R) < 5 ~{\rm GeV}~ (< 0.1 p^\ell_T~{\rm for}~ PE)
$$

\item{iv)} The coverage in the aximuthal angle $\phi$ is 84 and 89\% for
central and plug electrons and 85\% for the muons.

\item{v)} The identification efficiency in 88\% (99\%) for the 1st (2nd)
central electron and 79\% for the plug electron, while it is 98\% for each
muon.

\item{vi)} The triggering efficiency is 98\% for $CE$ (91\% for $CM$); which is
required to cover at least one of the electrons (muons). The net
triggering efficiency for the dilepton events corresponds to
$$
f_{tr} = 1 - (1 - f^1_{tr}) (1-f^2_{tr}).
\eqno (13)
$$
Note that in each of the earlier cases the efficiency factor for the
dilepton events corresponds to the product of those for the individual
leptons. The cuts (i-iii) are incorporated in the Monte Carlo program; and
the resulting cross-section is multiplied by a combined efficiency factor
$$f (CE-CE, CE-PE, CM-CM, CM-MI) = .62,.51,.68,.63
\eqno (14)
$$
arising from (iv-vi). Finally a dilepton mass cut $M_{\ell\ell} \not=$
75-105 GeV is imposed to suppress the $Z$ decay background.

The resulting $\mu\mu$ and $ee$ cross-sections are shown against the
dilepton azimuthal angle in Fig. 1 for the exotic quark case. The
corresponding $CDF$ events are also shown for comparison. The $CDF$ events
are largely concentrated in the back-to-back direction [17], as expected
for the Drell-Yan and the residual $Z$ decay backgrounds. In contrast the
predicted cross-section are either isotropic or peaked at smaller
azimuthal opening angles depending on the heavy quark mass. The
corresponding distributions for the 4th generation quark case are very
similar. It is clear from Fig. 1 that an azimuthal cut of $\phi_{\ell\ell} <
120^\circ$ removes all the $\mu\mu$ events and most of the $ee$ events without
reducing the signal cross-section seriously. It may be mentioned here that
the corresponding azimuthal cut for the $t\bar t$ search [1] was at $160^\circ$
since most of the dilepton events could be eliminated by a missing-$p_T$
cut. Since there is no neutrino and hence no missing-$p_T$ for the $FCNC$
decay, one has to impose a relatively stronger azimuthal cut.

Fig. 2 shows the predicted cross-sections after the $\phi_{\ell\ell} <
120^\circ$ cut, as functions of the heavy quark mass for (a) $\mu\mu$, (b)
$ee$, (c) $\mu\mu + ee$ channels. The predictions of the exotic and the
4th generation quark models are shown by solid and dashed lines
respectively. The right hand scale shows the corresponding number of
events for a common integrated luminosity of $4.1 ~pb^{-1}$; the $\mu\mu$
cross-section has been accordingly scaled down by a factor of 3.5/4.1. The
arrows indicate the 95\% $CL$ upper limits of 3 and 9.15 events
corresponding respectively to 0 $\mu\mu$ and 4 $ee$ events in the data.
Evidentry the strongest mass limits come from the $\mu\mu$ case (Fig. 2a);
i.e.
$$
m_Q ({\rm exotic,~4th~ gen.}) > 90, 80 ~{\rm GeV}
\eqno (15)
$$
at the 95\% $CL$. One sees from Fig. 2b that, while the $ee$ cross-section
is marginally larger than $\mu\mu$ the experimental limit is thrice as
large. Consequently the mass limits drop to 83 and 65 GeV for the two
models. For the same reason the combined $\mu\mu$ and $ee$ data, shown in
Fig. 2c, gives marginally lower mass limits than the $\mu\mu$ case -- i.e.
$m_Q = 88$ and 75 GeV for the exotic and 4th generation quarks
respectively. It should be noted here that the relative value of the two
mass limits essentially reflect the relative size of the branching ratios
for the tree and loop level $FCNC$ decay (8), which is 3:1. Indeed, a
comparison of the solid and dashed lines of Fig. 2 shows that the relative
size of the corresponding cross-sections is paractically identical to this
ratio -- i.e. the final cross-section is insensitive to the detailed shape
of the decay matrix element, as mentioned earlier.

Although our program does not include jet fragmentation and jet energy
resolution, the only effect of jets on the dilepton cross-section is
through the lepton isolation cut. Since the isolation cut is rather mild
for a quark mass of 80-90 GeV (accounting for a loss of only $\sim$ 30\%
dilepton events), we do not expect the above effects to influence the
dilepton corss-section appreciably. We believe the largest sources of
uncertainty are the $QCD$ parametrisation alongwith the higher order $QCD$
correction. As mentioned earlier, they can increase or decrease the
cross-section by a factor of 1.5. The resulting uncertainty in the mass
limits of eq. (15) is $\pm 5$ GeV. Thus one may take the conservative mass
limits for exotic and 4th generation quarks to be $m_Q > 85$ and 75 GeV
respectively. Of course there is an additional source of uncertainty in
the latter case -- i.e. the $m_t$ and $m_{t'}$ dependence of the
corresponding branching ratio for the decay process (8). Having chosen a
conservative value for this branching ratio, however, we feel the above
conservative mass limit should not be affected.

Finally, let us compare our mass limits for quarks decaying via $FCNC$
with that obtained earlier for quarks decaying via $CC$ [1]. For exotic
quarks, having tree level $FCNC$ decay, the $\mu\mu$ branching ratio is
3.3\%. This means a 6.6\% branching ratio for the $Q\bar Q$ to decay into
the dimuon channel. This is already larger than the 4\% branching ratio
for $Q\bar Q$ to decay into all the dilepton channels in the $CC$ case.
While the lepton detection efficiencies for the two cases are similar,
the larger branching ratio for the former case is effectively compensated
by the stronger
$\phi_{\mu\mu}$ cut. Consequently one gets very similar mass limits for
the two cases. One can combine the two to obtain a conservative mass limit
of 85 GeV for exotic quarks, which is valid for all values of the quark
mixing angles. Similarly one can combine the mass limits for $CC$ and loop
level $FCNC$ decays to obtain a corresponding mass limit of 75 GeV for the
4th generation quark.

In summary, $FCNC$ decay occurs at tree level for exotic quarks (singlet,
vector doublet and mirror doublet) and at one-loop level for quarks of the
4th generation. For some of the heavy quarks, this is expected to dominate
over the $CC$ decay over a wide range of quark mixing angles in the first
case and a more limited range in the second. The present Tevatron mass
limit on heavy quarks would not apply in these cases, since it has been
obtained under the assumption of $CC$ decay. However, a systematic
analysis of the Tevatron $\mu\mu$ and $ee$ data gives comparable mass
limits for heavy quarks decaying via $FCNC$ - i.e. $m_Q > 85$ and 75 GeV
for the exotic and the 4th generation cases respectively. Taken together
with the earlier limit, it implies that these mass limits for exotic and
4th generation quarks are valid for all values of the quark mixing angles.

We thank N.K. Mondal for discussions regarding the Tevatron data.

\endpage

\underbar{\bf References and Foot Notes}

\item{[1]} CDF Collaboration: F. Abe et al., Phys. Rev. D45, 3921 (1992).
\item{[2]} CDF Collaboration: F. Abe et al., Phys. Rev. Lett. 64, 147 (1990).
\item{[3]} P. Langacker and D. London, Phys. Rev. D38, 886 (1988).
\item{[4]} V. Barger, N.G. Deshpande, R.J.N. Phillips and K. Whisnant,
Phys. Rev. D33, 1912 (1986); J. Rosner, Comments Nucl. Part. Phys. 15, 195
(1986); J.L. Hewett and T.L. Rizzo, Phys. Rep. 183, 193 (1989).
\item{[5]} F. del Aguila et al., Nucl. Phys. B334, 1 (1990).
\item{[6]} F. del Aguila, Ann. Phys. (NY) 165, 237 (1985); S. Singh and
N.K. Sharma, Phys. Rev. D36, 160 (1987).
\item{[7]} W.S. Hou and R.G. Stuart, Phys. Rev. Lett. 62, 617 (1989);
Nucl. Phys. B 320, 277 (1989) and B 349, 91 (1991).
\item{[8]} P. Agrawal, S.D. Ellis and W.S. Hou, Phys. Lett. 256 B, 289 (1991).
\item{[9]} P. Agrawal and W.S. Hou, Preprint PSI-PR-92-02 (1992).
\item{[10]} G. Bhattacharyya, A. Raychaudhuri, A. Datta and S.N. Ganguli;
Mod. Phys. Lett. A6, 2921 (1991); E. Nardi, E. Roulet and D. Tommasini,
Preprint Fermilab Pub-91/207-A(1991).
\item{[11]} ALEPH Collaboration: D. Decamp et al., Phys. Lett. 236B, 511
(1990); OPAL Collaboration: M.J. Akrawy et al., Phys. Lett 236B, 364
(1990) and 246B, 285 (1990).
\item{[12]} M. Gluck, F. Hoffmann and E. Reya, Z. Phys. C13, 119 (1982).
\item{[13]} M. Diemoz, F. Ferroni, E. Longo and G. Martinelli, Z. Phys.
C39, 21 (1988). We have used a simple analytic parametrisation of these
structure functions give by M. Gluck, R.M. Godbole and E. Reya, Dortmund
Preprint, DO-TH-89/16 (1989).
\item{[14]} G. Altarelli et al., Nucl. Phys. B308, 724 (1988); R.K. Ellis,
Phys. Lett. B259, 492 (1991).
\item{[15]} G.'t Hooft and M. Veltman, Nucl. Phys. B153, 365 (1979); G.
Pasarino and M. Veltman, Nucl. Phys. B160, 151 (1979).
\item{[16]} We will be happy to communicate them as well as the code for
the relevant Spence functions to any one interested in evaluating these
quantities.   The code for the spence function was written by Biswarup
Mukhopadhyaya and Amitava Raychoudhuri, Phys. Rev. D39, 280 (1989).
\item{[17]} The last bin of the $ee$ data contains about 40 events.

\endpage

\underbar{\bf Figure Captions}

\item{\rm Fig.~1} The distribution of the (a) $\mu\mu$ and (b) $ee$
cross-section in the azimuthal opening angle of the lepton pair for
different values of the heavy quark mass $m_Q$ (in GeV). The right hand
scale shows the corresponding number of events per $30^\circ$ for the
integrated luminosity of 3.5 and 4.1 $pb^{-1}$ for the $\mu\mu$ and $ee$
data [1]. The data points are shown by the histograms.

\item{\rm Fig.~2} The dilepton cross-sections after the $\phi_{\ell\ell} <
120^\circ$ cut are shown against the heavy quark mass for tree (solid
lines) and loop level (dashed lines) $FCNC$ decay. The right hand scale
shows the corresponding number of events for a luminosity of 4.1 $pb^{-1}$.
The arrows indicate the 95\% $CL$ upper limits of 3~$\mu\mu$ and 9.15 $ee$
events corresponding to 0 $\mu\mu$ and 4 $ee$ events in the data [1].

\bye